\documentstyle[12pt]{article}
\newcommand\be{\begin{equation}}
\newcommand\ee{\end{equation}}
\newcommand\bea{\begin{eqnarray}}
\newcommand\eea{\end{eqnarray}}
\newcommand\ket[1]{|#1\rangle}
\newcommand\bra[1]{\langle #1|}

\newcommand{\fatalpha}{{\bf \alpha \kern -0.44em \alpha}}
\newcommand{\fatsigma}{{\bf \sigma \kern -0.54em \sigma}}
\newcommand{\tpchi}{{\bf \chi \kern -0.35em \chi}}
\newcommand{\llambda}{{\bf \lambda \kern -0.45em \lambda}}

\bibliography{plain}
\title{\bf Best separable
approximation with semi-definite programming method} \vspace{20mm}
\author{ M. A. Jafarizadeh$^{a,b,c}$
 \thanks{E-mail:jafarizadeh@tabrizu.ac.ir}, M.Mirzaee$^{a,b}$
 \thanks{E-mail:mirzaee@tabrizu.ac.ir},
 M.Rezaee$^{a,b}$ \thanks{E-mail:karamaty@tabrizu.ac.ir}
\\
\\
$^a${\small Department of Theoretical Physics and Astrophysics,
Tabriz University, Tabriz 51664, Iran.} \\ $^b${\small Institute
for Studies in Theoretical Physics and Mathematics, Tehran
19395-1795, Iran.} \\ $^c${\small Research Institute for
Fundamental Sciences, Tabriz 51664, Iran.}} \pagebreak

\begin{document}
\maketitle \vspace{15mm}
\newpage
\begin{abstract}

The  present methods for obtaining the optimal Lewenestein-
Sanpera decomposition of a mixed state are difficult to handle
analytically. We provide a simple analytical expression for the
optimal Lewenstein-Sanpera decomposition by using semidefinite
programming.  Specially, we obtain the optimal Lewenstein-Sanpera
decomposition for some examples such as: Bell decomposable state,
Iso-concurrence state, generic two qubit state in Wootters's
basis, $2\otimes 3$ Bell decomposable state, $d\otimes d$ Werner
and isotropic states, a one parameter $3\otimes 3$ state and
finally multi partite
 isotropic state.

{\bf PACs Index: 03.65.Ud }
\end{abstract}
\newpage
\vspace{70mm}
\section{INTRODUCTION}
 Entanglement is one of the most striking features
of quantum mechanics \cite{EPR,shcro}.   In the case of pure
states it is easy to check whether a given state is, or is not
entangled.  For mixed states, however, the statistical properties
of the mixture can hide the quantum correlations embodied in the
system, making thus the distinction between separable and
entangled states enormously difficult.

In the pioneering paper \cite{LS}, a very interesting description
of entanglement was achieved by defining the best separable
approximation (BSA) of a mixed state. In the case of 2-qubit
system, it consists of a decomposition of the state into a linear
combination of mixed separable part and a pure entangled one. In
this way, the whole non-separability properties are concentrated
in the pure part.

In the Ref. \cite{LS}, the numerical method for finding the BSA
has been reported. Some analytical results are also obtained for
special states of two qubit states \cite{englert}. Further in
\cite{Wellens} the BSA of two qubit state has been obtained
algebraically. They have also shown that in some cases the weight
of the entangled part in the decomposition is equal to the
concurrence of the state. An attempt to generalize the results of
Ref \cite{LS} is made in \cite{karnas}.

There is another method we can use which achieves exactly the
same effect, called semidefinite programming(SDP). Over the past
years, SDP in particular, have come to be recognized as valuable
numerical tools for control system analysis and design. In SDP
one minimizes a linear function subject to the constraint that an
affine combination of symmetric matrices is positive
semidefinite. SDP, has been studied (under various names) as far
back as the 1940s. Since 1990 many applications have been
discovered in areas such as estimation, signal processing and it
is currently considered to be the hottest area in optimization.
Although SDP is designed to be applied in numerical methods it
can be used for analytical computations. All of the above
mentioned applications indicate, that the method of SDP is very
useful. Some authors try to use the SDP to construct an explicit
entanglement witness and entanglement
distillation\cite{RAIN,VERS,DOHER}.

In this paper we use the   SDP method in order to obtain the
optimal Lewenstein-Sanpera decomposition  (LSD) of a mixed state.
Then we show how to perform the optimal LSD for well known
different examples via SDP method.

The paper is organized as follows:\\ In section-2 we define SDP.
In section -3 we give a brief review of
 Optimal LSD. In section -4, using the SDP method we obtain the optimal
 LSD of some mixed state density matrices such as: Bell decomposable state, Iso-concurrence
state, generic two qubit state in Wootters's basis, $2\otimes 3$
Bell decomposable state, $d\otimes d$ Werner and isotropic
states, a one parameter $3\otimes 3$ state and finally multi
partite  isotropic state  . The paper is ended with a brief
conclusion.

\section{Semi-definite programming}\label{semi}
A SDP  is a particular type of convex optimization problem
\cite{optimize}. A SDP problem requires minimizing a linear
function subject to a linear matrix inequality (LMI) constraint
\cite{optimize1}: \be \label{sdp1}\begin{array}{cc}
\mbox{minimize} & {\cal P}=c^{T}x
\\ \mbox{subject to} & F(x)\geq 0,
\end{array}\ee where c is a given vector,
$x^{T}=(x_{1},...,x_{n}), $ and $F(x)=F_{0}+\sum_{i}x_{i}F_{i},$
for some fixed hermitian matrices $F_{i}$. The inequality sign in
$F(x)\geq 0$ means that $F(x)$ is positive semidefinite.

This problem is called the primal problem. Vectors x whose
components are the variables of the problem and satisfy the
constraint $F(x) \geq 0$ are called primal feasible points, and
if they satisfy $F(x) > 0$ they are called strictly feasible
points. The minimal objective value $c^{T} x$ is by convention
denoted by ${\cal P}^{\ast}$ and is called the primal optimal
value.

Due to the convexity of  set of feasible points, SDP  has a nice
duality structure, with, the associated dual program being: \be
\label{sdp2}\begin{array}{cc} \mbox{maximize} & -Tr[F_{0}Z] \\
    & Z\geq 0 \\  & Tr[F_{i}Z]=c_{i}. \end{array}\ee

Here the variable is the real symmetric (or Hermitean) matrix Z,
and the data c, $F_{i}$ are the same as in the primal problem.
Correspondingly, matrices Z satisfying the constraints are called
dual feasible (or strictly dual feasible if $Z > 0$). The maximal
objective value of $-Tr F_{0}Z$, i.e., the dual optimal value, is
denoted by $d^{\ast}$.

The objective value of a primal(dual) feasible point is an upper
(lower) bound on ${\cal P}^{\ast}$($d^{\ast}$.  The main reason
why one is interested in the dual problem is that one can prove
that $d^{\ast} \leq {\cal P}^{\ast}$, and under relatively mild
assumptions, we can have ${\cal P}^{\ast} = d^{\ast}$. If the
equality holds, one can prove the following optimality condition
on $x$:

A primal feasible $x$ and a dual feasible $Z$ are optimal which is
denoted by $\hat{x}$ and $\hat{Z}$ if and only if  \be
\label{slacknes} F(\hat{x}) \hat{Z}=\hat{Z} F(\hat{x})=0. \ee This
latter condition is called the complementary slackness condition.

In one way or another, numerical methods for solving SDP problems
always exploit the inequality $d \leq d^{\ast} \leq {\cal
P}^{\ast} \leq {\cal P}$, where d and ${\cal P}$ are the objective
values for any dual feasible point and primal feasible point,
respectively. The difference \be\label{sdp3}{\cal
P}^{\ast}-d^{\ast}= c^{T}x+ Tr[F_{0}Z]= Tr[F(x)Z]\geq 0 \ee is
called the duality gap. If the equality $d^{\ast}={\cal
P}^{\ast}$ holds, i.e., the optimal duality gap is zero, then we
say that strong duality holds.
\section{Optimal Lewenstein-Sanpera
decomposition}\setcounter{equation}{0}
 According to pioneering work of
Lewenstein and Sanpera \cite{LS}, any bipartite density matrix
$\rho$ has a decomposition of the form \be \rho=(1-\lambda)
\rho_{e}+\lambda\rho^{'}_{s},\ee where $\rho^{'}_{s}$ is a
separable density matrix, $\rho_{e}$ is a entangled state, and
the parameter $\lambda\in [0 , 1]$. According to the theorem (2)
of reference \cite{LS},   any $2-$qubit density matrix $\rho$ can
be written as \be
\rho=\lambda\rho^{'}_{s}+(1-\lambda)\ket{\psi}\bra{\psi}\ee where
the entangled part is pure state, but in general $\rho_{e}$ can be
a mixed or a pure state where the whole  entanglement of $\rho$ is
concentrated in $(1-\lambda) \rho_{e}$.

 Rare
exceptions aside, the LSD of a given (non separable) $\rho$ is
not unique, there is usually a continuum of LSD  to choose from.
The decomposition with the largest weight $\lambda$ of the
separable part is the optimal LSD with respect to the chosen
separable set, which is proved to be uniquely determined.
According to its definition, the separable part of this
decomposition is called the best separable approximation (BSA) of
$\rho$, and its weight $\lambda$ the separability.
\section{Optimal Lewenstein-Sanpera
decomposition with semi-definite
programming}\setcounter{equation}{0} In this section  using the
SDP method we obtain the optimal LSD of a mixed state for some
well known different mixed states.
\subsection{Optimal Lewenstein-Sanpera
decomposition for Bell-decomposable state} A Bell decomposable
(BD) state is defined by:
\begin{equation}
\rho=\sum_{i=1}^{4}p_{i}\left|\psi_i\right>\left<\psi_i\right|,\quad\quad
0\leq p_i\leq 1,\quad \sum_{i=1}^{4}p_i=1,
 \label{BDS1}
\end{equation}
where $\left|\psi_i\right>$ is Bell state, given by:
\begin{eqnarray}
\label{BS1} \left|\psi_1\right>=\left|\phi^{+}\right>
=\frac{1}{\sqrt{2}}(\left|\uparrow\uparrow\right>
+\left|\downarrow\downarrow\right>), \\
\label{BS2}\left|\psi_2\right>=\left|\phi^{-}\right>
=\frac{1}{\sqrt{2}}(\left|\uparrow\uparrow\right>
-\left|\downarrow\downarrow\right>), \\
\label{BS3}\left|\psi_3\right>=\left|\psi^{+}\right>
=\frac{1}{\sqrt{2}}(\left|\uparrow\downarrow\right>
+\left|\downarrow\uparrow\right>), \\
\label{BS4}\left|\psi_4\right>=\left|\psi^{-}\right>
=\frac{1}{\sqrt{2}}(\left|\uparrow\downarrow\right>
-\left|\downarrow\uparrow\right>).
\end{eqnarray}
In terms of Pauli's matrices, $\rho$ can be written as,

\begin{equation}
\rho=\frac{1}{4}(I\otimes I+\sum_{i=1}^{3}
t_i\sigma_{i}\otimes\sigma_{i}), \label{BDS2}
\end{equation}
where \cite{Horod}

\begin{equation}\label{t-p}
\begin{array}{rl}
t_1=&p_1-p_2+p_3-p_4,  \\
t_2=&-p_1+p_2+p_3-p_4, \\
t_3=&p_1+p_2-p_3-p_4.
\end{array}
\end{equation}
From the positivity of $\rho$ we get
\begin{equation}\label{T1}
\begin{array}{rl}
1+t_1-t_2+t_3\geq & 0,  \\
1-t_1+t_2+t_3\geq & 0,  \\
1+t_1+t_2-t_3\geq & 0,  \\
1-t_1-t_2-t_3\geq & 0.
\end{array}
\end{equation}
These equations form a tetrahedral  with its vertices located at
$(1,-1,1)$, $(-1,1,1)$, $(1,1,-1)$, $(-1,-1,-1)$ \cite{Horod} .
In fact these vertices denote the Bell states given in Eqs.
(\ref{BS1}) to (\ref{BS4}), respectively.

On the other hand  $\rho$ given in Eq. (\ref{BDS2}) is separable
if and only if $t_i$ satisfy Eq. (\ref{T1}) together with the
following equation
\begin{equation}\label{T2}
\begin{array}{rl}
1+t_1+t_2+t_3\geq & 0,  \\
1-t_1-t_2+t_3\geq & 0,  \\
1+t_1-t_2-t_3\geq & 0,  \\
1-t_1+t_2-t_3\geq & 0.
\end{array}
\end{equation}

Inequalities (\ref{T1}) and (\ref{T2}) form an octahedral with its
 vertices located at
$O_1^{\pm}=(\pm 1,0,0)$, $O_2^{\pm}=(0,\pm 1,0)$ and
$O_3^{\pm}=(0,0,\pm 1)$. So, tetrahedral is divided into five
regions. Central regions, defined by octahedral, are separable
states ($p_{k}\leq \frac{1}{2}$). There are also four smaller
equivalent tetrahedral corresponding to entangled states($p_{k}>
\frac{1}{2}$ for only one of $k=1,...,4$), where
$p_{k}=\frac{1}{2}$ denote to boundary between separable and
entangled region. Each tetrahedral takes one Bell state as one of
its vertices.

Now in order to obtain optimal LSD of entangled BD state given in
(\ref{BDS1}), with $p_{1}>\frac{1}{2}$, we first choose an
arbitrary separable state  \be\label{ICDS3}
\rho^{'}_{s}=\sum_{i=1}^{4}p^{'}_{i}\left|\phi_i\right>\left<\phi_i\right|,\quad\quad
0\leq p_i\leq
1,\;\;\sum_{i=1}^{4}p^{'}_i=1,\;\;\;p^{'}_1<\frac{1}{2}\;,\;\sum_{i=1}^{4}p^{'}_{i}=1,\ee
in the separable region. Then using strict  SDP  optimization
prescription of section (2),  we try to optimize
$Tr(\Lambda\rho_{s}^{\prime})$ with respect to
$\rho-\Lambda\rho_{s}^{\prime} > 0$, where the feasible solution
corresponds to \be \label{ICDS4}
\Lambda_{max}=min\{\frac{p_{1}}{p_{1}^{\prime}},\frac{p_{2}}{p_{2}^{'}},\frac{p_{3}}{p_{3}^{'}},\frac{p_{4}}{p_{4}^{'}}
\}. \ee Now, using the inequalities \be\label{BD7}
\Lambda_{max}\leq \frac{p_{i}}{p^{'}_{i}},\;\; for\;i=2,3,4\ee
 and summing over the indices  i=2,3 and 4, we obtain
\be (1-p^{'}_{1})\Lambda_{max} \leq (1-p_{1}),\ee since, we have
\be\frac{p_{1}}{p_{1}^{\prime}} \geq
\frac{1-p_{1}}{1-p^{'}_{1}}\;\;,p_{1} >\frac{1}{2} \geq
p_{1}^{\prime}. \ee The only possible choice of $\Lambda_{max}$
which is  consistent with the positivity of $\rho
-\Lambda_{max}\rho_{s}^{\prime}$ is \be \label{BD8}
\Lambda_{max}= \frac{(1-p_{1})}{(1-p^{'}_{1})}.\ee This choice of
$\Lambda_{max}$ given in (\ref{BD8}) saturates the inequalities
(\ref{BD7}) and turns the inequalities to equalities, that is, we
have $p^{\prime}_{i}=\frac{p_{i}}{\Lambda_{max}}\;,i=2,3,4$. The
equation (\ref{BD8}) indicates that $\Lambda_{max}$ is a
monotonic increasing function of $p^{\prime}_{1}$ and its maximum
value corresponds to $p_{1}^{\prime}=\frac{1}{2}$, with \be
\Lambda_{max}=2(1-p_{1}).\ee and \be
p^{\prime}_{i}=\frac{p_{i}}{2(1-p_{1})}\;,i=2,3,4.\ee Substituting
the results that obtained for $\Lambda_{max}$ and
$p_{i}^{\prime}\;,i=1,2,3,4 $ in
$\rho-\Lambda_{max}\rho_{s}^{\prime}$, we obtain
\be\rho-\Lambda_{max}\rho^{'}_{s}=(2p_{1}-1)\ket{\phi_{1}}\bra{\phi_{1}},\ee
which is a pure state in agreement with theorem (2) of
Ref.\cite{LS}.

Therefore, we have
\be\rho=2(1-p_{1})\rho^{'}_{s}+(2p_{1}-1)\ket{\phi_{1}}\bra{\phi_{1}},\ee
which is optimal LSD of BD states in agreement with
\cite{Wellens}.
\subsection{Iso-concurrence decomposable states}\label{subsecICD}
In this section we define iso-concurrence decomposable (ICD)
states, then we obtain optimal LSD for this example. The
iso-concurrence states are defined by
\cite{akhtar2,akhtar1,Ericsson}
\begin{eqnarray} \label{ICS12}
\left|\phi_1\right>=\cos{\theta}\left|00\right>
+\sin{\theta}\left|11\right>),\qquad
\left|\phi_2\right>=\sin{\theta}\left|00\right>
-\cos{\theta}\left|11\right>), \\
\label{ICS34} \left|\phi_3\right>=\cos{\theta}\left|01\right>
+\sin{\theta}\left|10\right>),\qquad
\left|\phi_4\right>=\sin{\theta}\left|01\right>
-\cos{\theta}\left|10\right>).
\end{eqnarray}
It is quite easy to see that the above states are orthogonal and
thus span the Hilbert space of $2\otimes 2$ systems. Now we can
define ICD states as
\begin{equation} \label{ICDS}
\rho=\sum_{i=1}^{4}p_{i}\left|\phi_i\right>\left<\phi_i\right|,\quad\quad
0\leq p_i\leq 1,\quad \sum_{i=1}^{4}p_i=1.
\end{equation}
 These states form a four simplex (tetrahedral)  with its
vertices defined by $p_1=1$, $p_2=1$, $p_3=1$ and $p_4=1$,
respectively.

Peres-Horodeckis criterion \cite{peres,horo0} for separability
implies that the state given in Eq. (\ref{ICDS}) is separable if
and only if the following inequalities are satisfied
\begin{eqnarray}
\label{ppt1} (p_1-p_2)\leq
\sqrt{4p_3p_4/\sin^2{2\theta}+(p_3-p_4)^2}, \\ \label{ppt2}
(p_2-p_1)\leq \sqrt{4p_3p_4/\sin^2{2\theta}+(p_3-p_4)^2}, \\
\label{ppt3} (p_3-p_4)\leq
\sqrt{4p_1p_2/\sin^2{2\theta}+(p_1-p_2)^2}, \\ \label{ppt4}
(p_4-p_3)\leq \sqrt{4p_1p_2/\sin^2{2\theta}+(p_1-p_2)^2}.
\end{eqnarray}
Inequalities (\ref{ppt1}) to (\ref{ppt4}) divide tetrahedral of
density matrices to five regions. The central regions, defined by
the above inequalities, form a deformed octahedral and are
separable states. In the other four  regions one of the above
inequality will not hold, therefore they represent entangled
states. Bellow we consider entangled states corresponding to the
violation of inequality (\ref{ppt1}) i.e. the states which
satisfy the following inequality
\begin{equation}
\label{ICDE1}
(p_1-p_2)>\sqrt{4p_3p_4/\sin^2{2\theta}+(p_3-p_4)^2}.
\end{equation}

Now in order to obtain optimal LSD of entangled Iso-concurrence
decomposable state given in (\ref{BDS1}), with
$(p_1-p_2)>\sqrt{4p_3p_4/\sin^2{2\theta}+(p_3-p_4)^2}$, we first
choose an arbitrary separable state \be\label{ICDS3}
\rho^{'}_{s}=\sum_{i=1}^{4}p^{'}_{i}\left|\phi_i\right>\left<\phi_i\right|,\quad\quad
0\leq p_i\leq
1,\;\;\sum_{i=1}^{4}p^{'}_i=1,\;\;\;p^{'}_1<p^{'}_2+\sqrt{4p^{'}_3p^{'}_4/\sin^2{2\theta}+(p^{'}_3-p^{'}_4)^2}.\ee
in the separable region. Then using strict  SDP  optimization
prescription of section (2),  we try to optimize
$Tr(\Lambda\rho_{s}^{\prime})$ with respect to
$\rho-\Lambda\rho_{s}^{\prime} > 0$, where the feasible solution
corresponds to \be \label{ICDS4}
\Lambda_{max}=min\{\frac{p_{1}}{p_{1}^{\prime}},\frac{p_{2}}{p_{2}^{'}},\frac{p_{3}}{p_{3}^{'}},\frac{p_{4}}{p_{4}^{'}}
\}. \ee Now, using the inequalities \be\label{ICD7}
\Lambda_{max}\leq \frac{p_{i}}{p^{'}_{i}},\;\; for\;i=2,3,4\ee
 and summing over the indices  i=2,3 and 4, we obtain
\be (1-p^{'}_{1})\Lambda_{max} \leq (1-p_{1}),\ee since, we have
\be\frac{p_{1}}{p_{1}^{\prime}} \geq
\frac{1-p_{1}}{1-p^{'}_{1}}\;\;,p_1>p_2+\sqrt{4p_3p_4/\sin^2{2\theta}+(p_3-p_4)^2}\geq
p_{1}^{\prime}. \ee The only possible choice of $\Lambda_{max}$
which is  consistent with the positivity of $\rho
-\Lambda_{max}\rho_{s}^{\prime}$ is \be \label{ICD8}
\Lambda_{max}= \frac{(1-p_{1})}{(1-p^{'}_{1})}.\ee This choice of
$\Lambda_{max}$ given in (\ref{ICD8}) saturates the inequalities
(\ref{ICD7}) and turns the inequalities to equalities, that is, we
have $p^{\prime}_{i}=\frac{p_{i}}{\Lambda_{max}}\;,i=2,3,4$. The
equation (\ref{ICD8}) indicates that $\Lambda_{max}$ is a
monotonic increasing function of $p^{\prime}_{1}$ and its maximum
value corresponds to \be \label{ICD9}
p_{1}^{\prime}=p^{'}_2+\sqrt{4p^{'}_3p^{'}_4/\sin^2{2\theta}+(p^{'}_3-p^{'}_4)^2},\ee
with \be \Lambda_{max}=\frac{1-p_{1}}{1-p^{\prime}_{1}}.\ee and
\be
p^{\prime}_{i}=\frac{p_{i}(1-p^{\prime}_{1})}{(1-p_{1})}\;,i=2,3,4.\ee
According to relation
 (\ref{ICD9}) we have \be
\Lambda_{max}=1-(p_{1}-p_{2})+\sqrt{4p_3p_4/\sin^2{2\theta}+(p_3-p_4)^2}.\ee
Substituting the results that obtained for $\Lambda_{max}$ and
$p_{i}^{\prime}\;,i=1,2,3,4 $ in
$\rho-\Lambda_{max}\rho_{s}^{\prime}$, we obtain
\be\rho-\Lambda_{max}\rho^{'}_{s}=\frac{C}{\sin{2\theta}}\ket{\phi_{1}}\bra{\phi_{1}},\ee
( $C$ is concurrence defined in \cite{akhtar1}) which is a pure
state in agreement with theorem (2) of Ref.\cite{LS}.

Therefore, we have
\be\rho=(1-(p_{1}-p_{2})+\sqrt{4p_3p_4/\sin^2{2\theta}+(p_3-p_4)^2})\rho^{'}_{s}+(2p_{1}-1)\ket{\phi_{1}}\bra{\phi_{1}},\ee
which is optimal LSD of ICD states in agreement with
\cite{Wellens}.

In the special case of ($\theta=\pi/4$) we obtain Bell
decomposable state and $\Lambda_{max}=2(1-p_{1})$.

\subsection{A generic $2 \times 2 $ density matrix in Wootters's basis}
In this subsection we obtain optimal LSD for a generic two qubit
density matrix by using Wootters basis. Wootters in \cite{woot}
has shown that for any two qubit density matrix $\rho$ there
always exist a decomposition
\be\label{wootters1}\rho=\sum_{i}\ket{x_{i}}\bra{x_{i}}\ee called
Wootters's basis, such that \be \label{wootters}\langle x_{i } |
\tilde{x_{j}}\rangle =\lambda_{i}\delta_{ij}\ee where
$\lambda_{i}$ are square roots of eigenvalues, in decreasing
order, of the non-Hermitian matrix $\rho\tilde{\rho}$ and \be
\tilde{\rho}=(\sigma_{y}\otimes \sigma_{y})
\rho^{\ast}(\sigma_{y}\otimes \sigma_{y}) \ee where $\rho^{\ast}$
is the complex conjugate of $\rho$ when it is expressed in a
standard basis such as
$\{\ket{\uparrow\uparrow},\ket{\uparrow\downarrow},
\ket{\downarrow\uparrow}, \ket{\downarrow\downarrow} \}$ and
$\sigma_{y}$ represent Pauli matrix in local basis
$\{\ket{\uparrow} , \ket{\downarrow} \}$. Based on this, the
concurrence of the mixed state $\rho$ is defined by max(0,
$\lambda_{1}-\lambda_{2}-\lambda_{3}-\lambda_{4}$) \cite{woot}.

Now let us define states $\ket{x^{\prime}_{i}}$ as
\be\label{wooters3}
\ket{x_{i}}=\frac{\ket{x^{\prime}_{i}}}{\sqrt{\lambda_{i}}},\;\;\;
for\; i=1,2,3,4. \ee Then $\rho$ can be expanded as \be
\rho=\sum_{i} \lambda_{i}
\ket{x^{\prime}_{i}}\bra{x^{\prime}_{i}} \ee and Eq.
(\ref{wootters}) takes the following form \be \langle
x^{\prime}_{i } | \tilde{x^{\prime}_{j}}\rangle =\delta_{ij} \ee
Also Wootters has shown that  $\rho$ is separable if
$\lambda_{1}-\lambda_{2}-\lambda_{3}-\lambda_{4} \leq 0$ and if
$\rho$ is in boundary separable state then
$\lambda_{1}=\lambda_{2}+\lambda_{3}+\lambda_{4}$. By defining
$P_{i}=\lambda_{i} K_{i}$, where $k_{i}=\langle x^{\prime}_{i} |
x^{\prime}_{i} \rangle $, then normalization condition of $\rho$
leads to  \be\label{woot1} Tr(\rho)=\sum_{i=1}^{4} P_{i}=1.\ee
Now in order to obtain optimal LSD of entangled Wootters state
given in (\ref{wootters1}), (\ref{wootters}), with $\lambda_{1}
>\lambda_{2}+\lambda_{3}+\lambda_{4}$, we first choose an
arbitrary separable state \be \rho^{'}_{s}=\sum_{i=1}^{4}
\lambda^{'}_{i} \ket{x^{\prime}_{i}} \bra{x^{\prime}_{i}}
,\quad\quad 0\leq \lambda^{\prime}_i\leq
\frac{1}{k_{i}},\;\;\sum_{i=1}^{4}\lambda^{'}_i
k_i=1,\;\;\;\lambda^{\prime}_{1}<\lambda^{\prime}_{2}+\lambda^{\prime}_{3}+\lambda^{\prime}_{4}.\ee
in the separable region. Then using strict  SDP  optimization
prescription of section (2),  we try to optimize
$Tr(\Lambda\rho_{s}^{\prime})$ with respect to
$\rho-\Lambda\rho_{s}^{\prime} > 0$, where the feasible solution
corresponds to \be \Lambda_{max}=min\{
\frac{\lambda_{1}}{\lambda^{'}_{1}},\frac{\lambda_{2}}{\lambda^{'}_{2}},
\frac{\lambda_{3}}{\lambda^{'}_{3}},\frac{\lambda_{4}}{\lambda^{'}_{4}}
\}=min\{\frac{p_{1}}{p_{1}^{\prime}},\frac{p_{2}}{p_{2}^{'}},\frac{p_{3}}{p_{3}^{'}},\frac{p_{4}}{p_{4}^{'}}
\}\ee Now, using the inequalities \be\label{wootters2}
\Lambda_{max}\leq \frac{p_{i}}{p^{'}_{i}},\;\; for\;i=2,3,4\ee
 and summing over the indices  i=2,3 and 4, we obtain
\be (1-p^{'}_{1})\Lambda_{max} \leq (1-p_{1})\ee since, we have
\be\frac{\lambda_{1}}{\lambda^{\prime}} \geq
\frac{1-\lambda_{1}}{1-\lambda^{'}_{1}}\;\;,\lambda_1>\lambda^{\prime}_{2}+\lambda^{\prime}_{3}+\lambda^{\prime}_{4}\geq
\lambda_{1}^{\prime}. \ee The only possible choice of
$\Lambda_{max}$ which is consistent with the positivity of $\rho
-\Lambda_{max}\rho_{s}^{\prime}$ is \be \label{wootters3}
\Lambda_{max}=
\frac{(1-p_{1})}{(1-p^{'}_{1})}=\frac{(1-k_{1}\lambda_{1})}{(1-k_{1}\lambda^{'}_{1})}.\ee
This choice of $\Lambda_{max}$ given in (\ref{wootters3})
saturates the inequalities (\ref{wootters2}) and turns the
inequalities to equalities, that is, we have
$\lambda^{\prime}_{i}=\frac{\lambda_{i}}{\Lambda_{max}}\;,i=2,3,4$.
The equation (\ref{wootters3}) indicates that $\Lambda_{max}$ is a
monotonic increasing function of $\lambda^{\prime}_{1}$ and its
maximum value corresponds to \be \label{wootters4}
\lambda_{1}^{\prime}=\lambda^{\prime}_{2}+\lambda^{\prime}_{3}+\lambda^{\prime}_{4},\ee
with \be
\Lambda_{max}=\frac{1-\lambda_{1}k_{1}}{1-\lambda^{\prime}_{1}k_{1}},\ee
and \be
\lambda^{\prime}_{i}=\frac{\lambda_{i}(1-\lambda^{\prime}_{1}k_{1})}{1-\lambda_{1}k_{1}},\ee
\be
\lambda^{\prime}_{1}=\frac{(\lambda_{2}+\lambda_{3}+\lambda_{4})}{1-\lambda_{1}k_{1}}
.\ee Therefore, we can show that  \be
\Lambda_{max}=1-k_{1}(\lambda_{1}-\lambda_{2}-\lambda_{3}-\lambda_{4})=1-k_{1}C,\ee
where $C$ is  concurrence. Using $k_{i}=\langle x^{\prime}_{i} |
x^{\prime}_{i} \rangle $ and (\ref{wooters3}) we obtain
\be\label{wo} \Lambda_{max}=1-\frac{C}{\lambda_{1}}\langle
x_{1}|x_{1}\rangle.\ee Substituting the results that obtained for
$\Lambda_{max}$ and $\lambda_{i}^{\prime}\;,i=1,2,3,4 $ in
$\rho-\Lambda_{max}\rho_{s}^{\prime}$, we obtain
\be\rho-\Lambda_{max}\rho^{'}_{s}=C\ket{x_{1}^{'}}\bra{x_{1}^{'}}
,\ee  which is a pure states in agreement with theorem (2) of
Ref.\cite{LS} .

Therefore, we have \be\rho=(1-\frac{C}{\lambda_{1}}\langle
x_{1}|x_{1}\rangle)\rho^{'}_{s}+C\ket{x_{1}^{'}}\bra{x_{1}^{'}}
,\ee which is optimal LSD of  states  in agreement with
\cite{Wellens}.
\subsection{ $2\otimes 3$ Bell decomposable state}
\label{subsecBDS23} In this subsection we obtain optimal LSD  for
the  Bell decomposable states of $2\otimes 3$ quantum systems. A
Bell decomposable density matrix acting on $2\otimes3$ Hilbert
space can be defined by
\begin{equation} \label{BDS23}
\rho=\sum_{i=1}^{6}p_{i}\left|\psi_i\right>\left<\psi_i\right|,\quad\quad
0\leq p_i\leq 1,\quad \sum_{i=1}^{6}p_i=1,
\end{equation}
where $\left|\psi_i\right>$ are Bell states in $H^6\cong
H^2\otimes H^3$ Hilbert space, defined by:

$$
\left|\psi_1\right>=
\frac{1}{\sqrt{2}}(\left|11\right>+\left|22\right>), \qquad
\left|\psi_2\right>=
\frac{1}{\sqrt{2}}(\left|11\right>-\left|22\right>),
$$
\begin{equation}\label{BS123456}
\left|\psi_3\right>=
\frac{1}{\sqrt{2}}(\left|12\right>+\left|23\right>), \qquad
\left|\psi_4\right>=
\frac{1}{\sqrt{2}}(\left|12\right>-\left|23\right>),
\end{equation}
$$ \left|\psi_5\right>=
\frac{1}{\sqrt{2}}(\left|13\right>+\left|21\right>), \qquad
\left|\psi_6\right>=
\frac{1}{\sqrt{2}}(\left|13\right>-\left|21\right>). $$ It is
quite easy to see that the above states are orthogonal and hence
they can  span the Hilbert space of $2\otimes3$ systems. From
Peres-Horodeckis \cite{peres,horo0} criterion for separability we
deduce that the state given in Eq. (\ref{BDS23}) is separable if
and only if the following inequalities are satisfied
\begin{equation}\label{S1}
(p_1-p_2)^2\le(p_3+p_4)(p_5+p_6),
\end{equation}
\begin{equation}\label{S2}
(p_3-p_4)^2\le(p_5+p_6)(p_1+p_2),
\end{equation}
\begin{equation}\label{S3}
(p_5-p_6)^2\le(p_1+p_2)(p_3+p_4).
\end{equation}
In the following we always assume without loss of generality that
$p_1\ge p_2$, $p_3 \ge p_4$ and $p_5 \ge p_6$.

Again in order to obtain optimal LSD of entangled BD state given
in (\ref{BDS23}), with
$p_{1}>p_{2}+\sqrt{(p_{3}+p_{4})(p_{5}+p_{6})}$, we first choose
an arbitrary separable state \be\label{BDS231}
\rho^{'}_{s}=\sum_{i=1}^{6}p^{'}_{i}\left|\phi_i\right>\left<\phi_i\right|,\quad\quad
0\leq p^{\prime}_i\leq
1,\;\;\sum_{i=1}^{6}p^{'}_i=1,\;\;\;p^{'}_1<p_{2}^{\prime}+\sqrt{(p_{3}^{\prime}+p_{4}^{\prime})(p_{5}^{\prime}+p_{6}^{\prime})};,\ee
in the separable region. Then using strict  SDP  optimization
prescription of section (2), we try to optimize
$Tr(\Lambda\rho_{s}^{\prime})$ with respect to
$\rho-\Lambda\rho_{s}^{\prime} > 0$, where the feasible solution
corresponds to \be \label{BDS232}
\Lambda_{max}=min\{\frac{p_{1}}{p_{1}^{\prime}},\frac{p_{2}}{p_{2}^{'}},\frac{p_{3}}{p_{3}^{'}},\frac{p_{4}}{p_{4}^{'}},\frac{p_{5}}{p_{5}^{\prime}},\frac{p_{6}}{p_{6}^{\prime}}
\}. \ee Now, using the inequalities \be\label{BDS233}
\Lambda_{max}\leq \frac{p_{i}}{p^{'}_{i}},\;\; for\;i=2,...,6\ee
 and summing over the indices  i=2,... and 6, we obtain
\be (1-p^{'}_{1})\Lambda_{max} \leq (1-p_{1}),\ee since, we have
\be\frac{p_{1}}{p_{1}^{\prime}} \geq
\frac{1-p_{1}}{1-p^{'}_{1}}\;\;,p_{1} >
p_{2}+\sqrt{(p_{3}+p_{4})(p_{5}+p_{6})}\;\;, p_{1}^{\prime}<
p_{2}^{\prime}+\sqrt{(p_{3}^{\prime}+p_{4}^{\prime})(p_{5}^{\prime}+p_{6}^{\prime})}.
\ee The only possible choice of $\Lambda_{max}$ which is
consistent with the positivity of $\rho
-\Lambda_{max}\rho_{s}^{\prime}$ is \be \label{BDS234}
\Lambda_{max}= \frac{(1-p_{1})}{(1-p^{'}_{1})}.\ee This choice of
$\Lambda_{max}$ given in (\ref{BDS234}) saturates the
inequalities (\ref{BDS233}) and turns the inequalities to
equalities, that is, we have
$p^{\prime}_{i}=\frac{p_{i}}{\Lambda_{max}}\;,i=2,...,6$. The
equation (\ref{BDS234}) indicates that $\Lambda_{max}$ is a
monotonic increasing function of $p^{\prime}_{1}$ and its maximum
value corresponds to
$p_{1}^{\prime}=p_{2}^{\prime}+\sqrt{(p_{3}^{\prime}+p_{4}^{\prime})(p_{5}^{\prime}+p_{6}^{\prime})}$,
with \be
\Lambda_{max}=\frac{(1-p_{1})}{(1-p_{1}^{\prime})}=1-p_{1}+
p_{2}+\sqrt{(p_{3}+p_{4})(p_{5}+p_{6})}.\ee and \be
p^{\prime}_{i}=\frac{p_{i}}{\Lambda_{max}}\;,i=2,...,6.\ee
Substituting the results that obtained for $\Lambda_{max}$ and
$p_{i}^{\prime}\;,i=1,...,6 $ in
$\rho-\Lambda_{max}\rho_{s}^{\prime}$, we obtain
\be\rho-\Lambda_{max}\rho^{'}_{s}=(1-\Lambda_{max})\ket{\phi_{1}}\bra{\phi_{1}},\ee
which is a pure states in agreement with theorem (2) of
Ref.\cite{LS}.

Therefore, we have \be\rho=(1-p_{1}+
p_{2}+\sqrt{(p_{3}+p_{4})(p_{5}+p_{6})})\rho^{'}_{s}+(p_{1}-
p_{2}-\sqrt{(p_{3}+p_{4})(p_{5}+p_{6})})\ket{\phi_{1}}\bra{\phi_{1}},\ee
which is optimal LSD of $2\times 3$, BD states.

The above choice of $\Lambda_{max}$ do not cover the whole set of
separable states lying at boundary $p_{1}=
p_{2}+\sqrt{(p_{3}+p_{4})(p_{5}+p_{6})}$. Hence we should try
other possible values of $\Lambda_{max}$  as follows: \be
p^{'}_{i}\Lambda_{max}\leq p_{i}\;\;,\; i=1,4,5,6.\ee Summing over
indices  $i=1,2,...,4 $ we obtain  \be \Lambda_{max}\leq
\frac{1-p_{2}-p_{3}}{1-p_{2}^{'}-p_{3}^{'}}. \ee Therefore,
maximum feasible choice of $\Lambda_{max}$ is \be
\Lambda_{max}=\frac{1-p_{2}-p_{3}}{1-p_{2}^{'}-p_{3}^{'}}, \ee
which is possible if we  choose \be
p^{'}_{i}=\frac{p_{i}}{\Lambda_{max}}\;\;, i=1,4,5,6, \ee Now
substituting $p^{'}_{i}=\frac{p_{i}}{\Lambda_{max}}\;\;,
i=1,4,5,6$ in normalization condition
$Tr(\rho_{s}^{'})=\sum_{i=1}^{6}=1$ and the separability equation
(\ref{S1}) we can solve $\Lambda_{max}$ as  a function of single
variable $p^{'}_{3}$ and after optimizing it with respect to
$p^{'}_{3}$ we get \be
\Lambda_{max}=1-(p_{2}-p_{1})-(p_{3}+p_{4})-\frac{1}{4}(p_{5}+p_{6}),
\ee
 and
\be
p^{'}_{2}=\frac{2p_{1}-p_{5}-p_{6}}{2\Lambda_{max}}\;\;\;,p^{'}_{3}=\frac{p_{5}+p_{6}-4p_{4}}{4\Lambda_{max}}.
\ee Similarly, following  the above procedure for other possible
choices, such as \be \Lambda_{max}=\frac{p_{i}}{p^{'}_{i}}\;\;,
i=1,3,4,6, \ee yields \be
p^{'}_{2}=\frac{2p_{1}-p_{3}-p_{4}}{2\Lambda_{max}}\;\;\;,p^{'}_{5}=\frac{p_{3}+p_{4}-4p_{6}}{4\Lambda_{max}}.
\ee \be
\Lambda_{max}=1-(p_{2}-p_{1})-(p_{5}+p_{6})-\frac{1}{4}(p_{3}+p_{4}).
\ee Of course, using this procedure we can obtain an optimal
separate decomposition with rank-3 entangled part of some
particular given density matrices.

\subsection{Werner states}\label{subsecWerner}
The Werner states are the only states that are invariant under
local unitary operations. For $d\otimes d$ systems the Werner
states are defined by \cite{werner}
\begin{equation}\label{werner1}
\rho_f=\frac{1}{d^3-d}\left((d-f)I+(df-1)F\right), \qquad -1\le
f\le 1,
\end{equation}
where $I$ stands for identity operator and $F=\sum_{i,j}\left|i
j\right>\left<j i \right|$. It is shown that Werner state is
separable iff $0\le f\le 1$.

Now in order to obtain optimal LSD of entangled Werner state given
in (\ref{werner1}), with $-1 < f < 0$, we first choose an
arbitrary separable state
\begin{equation}\label{werner1}
\rho_{s}^{\prime}=\rho_f^{\prime}=\frac{1}{d^3-d}\left((d-f^{\prime})I+(df^{\prime}-1)F\right),
\qquad 0\le f^{\prime}\le 1,
\end{equation}
in the separable region. Then using strict  SDP  optimization
prescription of section (2),  we try to optimize
$Tr(\Lambda\rho_{s}^{\prime})$ with respect to
$\rho-\Lambda\rho_{s}^{\prime} > 0$, where the feasible solution
corresponds to \be \label{werner2}
\Lambda_{max}=min\{\frac{(f+1)}{(f^{\prime}+1)},\frac{(1-f)}{(1-f^{\prime})}
\}=\frac{(f+1)}{(f^{\prime}+1)}. \ee The equation (\ref{werner2})
indicates that $\Lambda_{max}$ is a monotonic increasing function
of $f^{\prime}$ and its maximum value corresponds to
$f^{\prime}=0$, with \be \Lambda_{max}=f+1.\ee Substituting the
results that obtained for $\Lambda_{max}$ in
$\rho-\Lambda_{max}\rho_{s}^{\prime}$, we obtain \be
\rho-\Lambda_{max}\rho^{'}_{s}=
\rho_{\{f\}}-\Lambda_{max}(1+f)\rho_{\{f=0\}}=f(\frac{F-I}{d^2-d}),\ee
which is a pure states in agreement with theorem (2) of
Ref.\cite{LS}.

Therefore, we have \be
\rho_{\{f\}}=(1+f)\rho_{\{f=0\}}+f(\frac{F-I}{d^2-d}).\ee which
is optimal LSD of Werner states.
\subsection{Isotropic states}\label{subsecIsotropic}
The $d\otimes d$ bipartite isotropic states are the only ones that
are invariant under $U\otimes U^\ast$ operations, where $^\ast$
denotes complex conjugation. The isotropic states of $d\otimes d$
systems are defined by \cite{horo3}
\begin{equation}\label{iso1}
\rho_F=\frac{1-F}{d^2-1}\left(I-\left|\psi^{+}\right>\left<\psi^{+}\right|\right)
+F\left|\psi^{+}\right>\left<\psi^{+}\right| , \qquad 0\le F\le 1,
\end{equation}
where $\left|\psi^{+}\right>=\frac{1}{\sqrt{d}}\sum_{i}\left|i i
\right>$ is maximally entangled state. It is shown that isotropic
state is separable when $0\le F\le \frac{1}{d}$ \cite{horo3}.

Now in order to obtain optimal LSD of entangled Isotropic state
given in (\ref{iso1}), with $\frac{1}{d} < f < 1$, we first
choose an arbitrary separable state
\begin{equation}\label{iso1}
\rho_{s}^{\prime}=\rho_F^{\prime}=\frac{1-F^{\prime}}{d^2-1}\left(I-\left|\psi^{+}\right>\left<\psi^{+}\right|\right)
+F^{\prime}\left|\psi^{+}\right>\left<\psi^{+}\right| , \qquad
0\le F^{\prime}\le \frac{1}{d},
\end{equation}
in the separable region. Then using strict  SDP  optimization
prescription of section (2), we try to optimize
$Tr(\Lambda\rho_{s}^{\prime})$ with respect to
$\rho-\Lambda\rho_{s}^{\prime} > 0$, where the feasible solution
corresponds to \be \label{werner2}
\Lambda_{max}=min\{\frac{(F)}{(F^{\prime})},\frac{(1-F)}{(1-F^{\prime})}
\}=\frac{(1-F)}{(1-F^{\prime})}. \ee The equation (\ref{werner2})
indicates that $\Lambda_{max}$ is a monotonic increasing function
of $F^{\prime}$ and its maximum value corresponds to
$F^{\prime}=\frac{1}{d}$, with \be
\Lambda_{max}=\frac{d(1-F)}{(d-1)}.\ee Substituting the results
that obtained for $\Lambda_{max}$ in
$\rho-\Lambda_{max}\rho_{s}^{\prime}$, we obtain \be
\rho_{F}-\Lambda_{max}\rho_{1/d}=(1-\Lambda_{max})\ket{\psi^{+}}\bra{\psi^{+}}\ee
which is a pure states in agreement with theorem (2) of
Ref.\cite{LS}.

Therefore, we have \be
\rho_{F}=\frac{(1-F)}{(1-F^{\prime})}\rho_{1/d}+(1-\Lambda_{max})\ket{\psi^{+}}\bra{\psi^{+}}\ee
which is optimal LSD of isotropic states.

\subsection{ One parameter $3\otimes 3$ state}\label{subsec33}
Let us consider a one parameter state acting on $H^9\cong
H^3\otimes H^3$ Hilbert space as \cite{horo2}
\begin{equation}\label{33}
\rho_{\alpha}=\frac{2}{7}\left|\psi^{+}\right>\left<\psi^{+}\right|+\frac{\alpha}{7}\sigma_{+}
+\frac{5-\alpha}{7}\sigma_{-}, \qquad 2\le \alpha \le 5,
\end{equation}
where
\begin{equation}
\begin{array}{l}
\left|\psi^{+}\right>=
\frac{1}{\sqrt{3}}\left(\left|11\right>+\left|22\right>+\left|33\right>\right),
\\
\sigma_{+}=\frac{1}{3}\left(\left|12\right>\left<12\right|
\left|23\right>\left<23\right|+\left|31\right>\left<31\right|\right),
\\
\sigma_{-}=\frac{1}{3}\left(\left|21\right>\left<21\right|
\left|32\right>\left<32\right|+\left|13\right>\left<13\right|\right).
\end{array}
\end{equation}
$\rho_{\alpha}$ is separable iff $2 \le \alpha \le 3$, it is bound
entangled iff $3 \le \alpha \le 4$
 and it is distillable entangled state iff $4
\le \alpha \le 5$ \cite{horo2}.

Now in order to obtain optimal LSD of entangled one parameter
$3\otimes 3$ state given in (\ref{33}), with $3 < \alpha < 5$, we
first choose an arbitrary separable state
\begin{equation}\label{331}
\rho_{s}^{\prime}=\rho_{\alpha^{\prime}}=\frac{2}{7}\left|\psi^{+}\right>\left<\psi^{+}\right|+\frac{\alpha^{\prime}}{7}\sigma_{+}
+\frac{5-\alpha^{\prime}}{7}\sigma_{-}, \qquad 2\le
\alpha^{\prime} \le 3,\end{equation} in the separable region.
Then using strict  SDP  optimization prescription of section (2),
 we try to optimize $Tr(\Lambda\rho_{s}^{\prime})$ with
respect to $\rho-\Lambda\rho_{s}^{\prime} > 0$, where the
feasible solution corresponds to \be \label{332}
\Lambda_{max}=\frac{5-\alpha}{5-\alpha^{\prime}}. \ee The equation
(\ref{332}) indicates that $\Lambda_{max}$ is a monotonic
increasing function of $\alpha^{\prime}$ and its maximum value
corresponds to $\alpha^{\prime}=3$, with \be
\Lambda_{max}=\frac{5-\alpha}{2}.\ee Substituting the results that
obtained for $\Lambda_{max}$ in
$\rho-\Lambda_{max}\rho_{s}^{\prime}$, we obtain \begin{equation}
\rho-\Lambda_{max}\rho_{\alpha=3}=(2/7\ket{\psi^{+}}\bra{\psi^{+}}+5/7\sigma_{+})\end{equation}
which is a pure states in agreement with theorem (2) of
Ref.\cite{LS}.

Therefore, we have \begin{equation}
\rho=(\frac{5-\alpha}{2})\rho_{\alpha=3}+(2/7\ket{\psi^{+}}\bra{\psi^{+}}+5/7\sigma_{+})\end{equation}
which is optimal LSD of one parameter $3\otimes 3$ states.

\subsection{ Multi partite isotropic states}\label{subsecMultiiso}
In this subsection we obtain the optimal LSD   for a n-partite
d-levels system. Let us consider the following mixture of the
completely random state $\rho_0=I/d^n$ and the maximally entangled
state $\left|\psi^{+}\right>$
\begin{equation}\label{multirho}
\rho(s)=(1-s)\frac{I}{d^n}+s\left|\psi^{+}\right>\left<\psi^{+}\right|,\qquad
0\le s \le 1,
\end{equation}
where $I$ denotes identity operator in $d^n$-dimensional Hilbert
space and
$\left|\psi^{+}\right>=\frac{1}{\sqrt{d}}\sum_{i=1}^{d}\left|ii\cdot\cdot\cdot
i\right>$. The separability properties of the state
(\ref{multirho}) is considered in Ref. \cite{pitt2}. It is shown
that the above state is separable iff $s \leq
s_0=\left(1+d^{n-1}\right)^{-1}$.

Now in order to obtain optimal LSD of entangled  multi partite
isotropic state given in (\ref{multirho}), with $s_{0} < s < 1$,
we first choose an arbitrary separable state
\begin{equation}\label{multirho1}
\rho_{s}^{\prime}=\rho(s^{\prime})=(1-s^{\prime})\frac{I}{d^n}+s^{\prime}\left|\psi^{+}\right>\left<\psi^{+}\right|,\qquad
0\le s^{\prime} \le s_{0},
\end{equation} in the separable region.
Then using strict  SDP  optimization prescription of section (2),
 we try to optimize $Tr(\Lambda\rho_{s}^{\prime})$ with
respect to $\rho-\Lambda\rho_{s}^{\prime} > 0$, where the
feasible solution corresponds to \be \label{multirho2}
\Lambda_{max}=\frac{1-s}{1-s^{\prime}}. \ee The equation
(\ref{multirho2}) indicates that $\Lambda_{max}$ is a monotonic
increasing function of $s^{\prime}$ and its maximum value
corresponds to $s^{\prime}=s_{0}$, with \be
\Lambda_{max}=\frac{1-s}{1-s_{0}}=\frac{(1-s)(1+d^{n-1})}{d^{n-1}}.\ee
Substituting the results that obtained for $\Lambda_{max}$ in
$\rho-\Lambda_{max}\rho_{s}^{\prime}$, we obtain \begin{equation}
\rho(s)-\Lambda_{max}\rho({s=s_{0}})=(1-\Lambda_{max})
\ket{\psi^{+}}\bra{\psi^{+}}\; \ee which is a pure states in
agreement with theorem (2) of Ref.\cite{LS}.

 Therefore, we have
\begin{equation}
\rho(s)=\frac{(1-s)(1+d^{n-1})}{d^{n-1}}\rho({s=s_{0}})+(1-\frac{(1-s)(1+d^{n-1})}{d^{n-1}})
\ket{\psi^{+}}\bra{\psi^{+}}\; \ee which is optimal LSD  of multi
partite isotropic states.

\section{Conclusion}
Here in this work we have been  able  to obtain LSD of bunch of
mixed state density matrices via an elegant method of convex
positive semidefinite optimization methods, where the results that
obtained   are  in agreement with those  obtained by other
methods in Ref.\cite{akhtar2,akhtar1}. Comparing this method with
those of previously introduced one, one can appreciate the
elegance and usefulness of SDP method in connection with  LSD.

\end{document}